\documentclass[aps,prl,reprint,superscriptaddress,showpacs]{revtex4-1}
\usepackage{graphicx}
\usepackage{dcolumn}
\usepackage{bm}

\begin{document}


\title{Radioactivity and Electron Acceleration in
Supernova Remnants}

\author{V.N.Zirakashvili}
\affiliation{Pushkov Institute of Terrestrial Magnetism, Ionosphere and Radiowave
Propagation,\\ 142190 Troitsk, Moscow Region, Russia }
\affiliation{Max-Planck-Institut f\"{u}r\ Kernphysik,  Saupfercheckweg 1, 69117
Heidelberg, Germany}
\author{F.A.Aharonian}
\affiliation{Dublin Institute for Advanced Studies, 31 Fitzwilliam Place, Dublin 2,
Ireland}
\affiliation{Max-Planck-Institut f\"{u}r\ Kernphysik,  Saupfercheckweg 1, 69117
Heidelberg, Germany}



\date{\today}

\begin{abstract}
We argue that the decays of radioactive nuclei related to $^{44}$Ti
and $^{56}$Ni ejected during supernova explosions  can provide a  vast
pool of mildly relativistic positrons and electrons  which are
further accelerated to ultrarelativistic energies
by reverse and forward  shocks.
This interesting  link between two independent processes -
the radioactivity and the particle acceleration -
can be a clue for solution of the well known theoretical problem
of electron injection  in supernova remnants. In the case of the brightest  radio source
Cas A, we demonstrate  that the radioactivity can supply adequate number of energetic
electrons and positrons  for interpretation of observational data provided that
they  are stochastically pre-accelerated in the upstream
regions of the  forward and reverse shocks.

\end{abstract}


\maketitle


\section{Introduction}

Supernova remnants (SNRs) are generally believed to be prime
candidates for production of both hadronic and electronic
components of galactic cosmic rays (CRs) via the diffusive shock
acceleration (DSA)  mechanism (see e.g. \cite{malkov01} for a
review). While the main aspects of the theory are well understood,
the key issue related to electrons is the so-called injection
problem which despite certain theoretical attempts (see e.g.
\cite{levinson94, bykov99}), remains an open question.  The
injection of  electrons is a serious challenge because the
electron gyroradius  is  small compared to   the shock
thickness which  is of the order of  the proton  gyroradius. In fact
this is a more general problem, related not only to DSA but also
to other electron acceleration mechanisms, e.g. through different
scenarios of stochastic acceleration \cite{fermiII}. In this paper
we  explore whether the  pool of suprathermal electrons and
positrons  related to the  decay products of radioactive nuclei
$^{56}$Ni and $^{44}$Ti can serve as an effective injector for
further acceleration of electrons in SNRs by the forward and  reverse shocks.

It is well established  that  the supernova ejecta contain
huge amount of radioactive nuclei. The decays of
these unstable nuclei have  been  proposed  as a source  of
low energy  positrons (see e.g. ref. \cite{chan93, martin10})
responsible for the  0.511 MeV annihilation
line observed from the direction of the Galactic Center.
In the case of the core-collapse supernova Cas~A,
approximately  0.1$M_{\odot}$ of $^{56}$Ni  has been  ejected
just after the explosion\cite{krause08}. The nuclei $^{56}$Ni decay with a
half lifetime  $t_{1/2}=6.1$~days into $^{56}$Co. Over the
first years after the explosion,  the decay products  of $^{56}$Co
($t_{1/2}=77$ days)  support the supernova optical light emission.
At later epochs, less abundant
radioactive nuclei with longer lifetimes  contribute to the
production of low energy suprathermal electrons, positrons and
gamma-rays. In particular, the detection of  characteristic
gamma-ray   \cite{iyudin94} and hard X-ray lines \cite{renaud06}
from $^{44}$Ti  gives a robust estimate of the total mass of
radioactive  $^{44}$Ti  ($t_{1/2}=63$ years) produced
in Cas~A:  $2\cdot 10^{-4}M_{\odot}$ . Recently a comparable amount
of $^{44}$Ti has been found also in the   youngest galactic supernova remnant -
SNR~G1.9+0.3 \cite{borkowski10}.

\begin{figure}[t]
\includegraphics[width=6.0cm,angle=270]{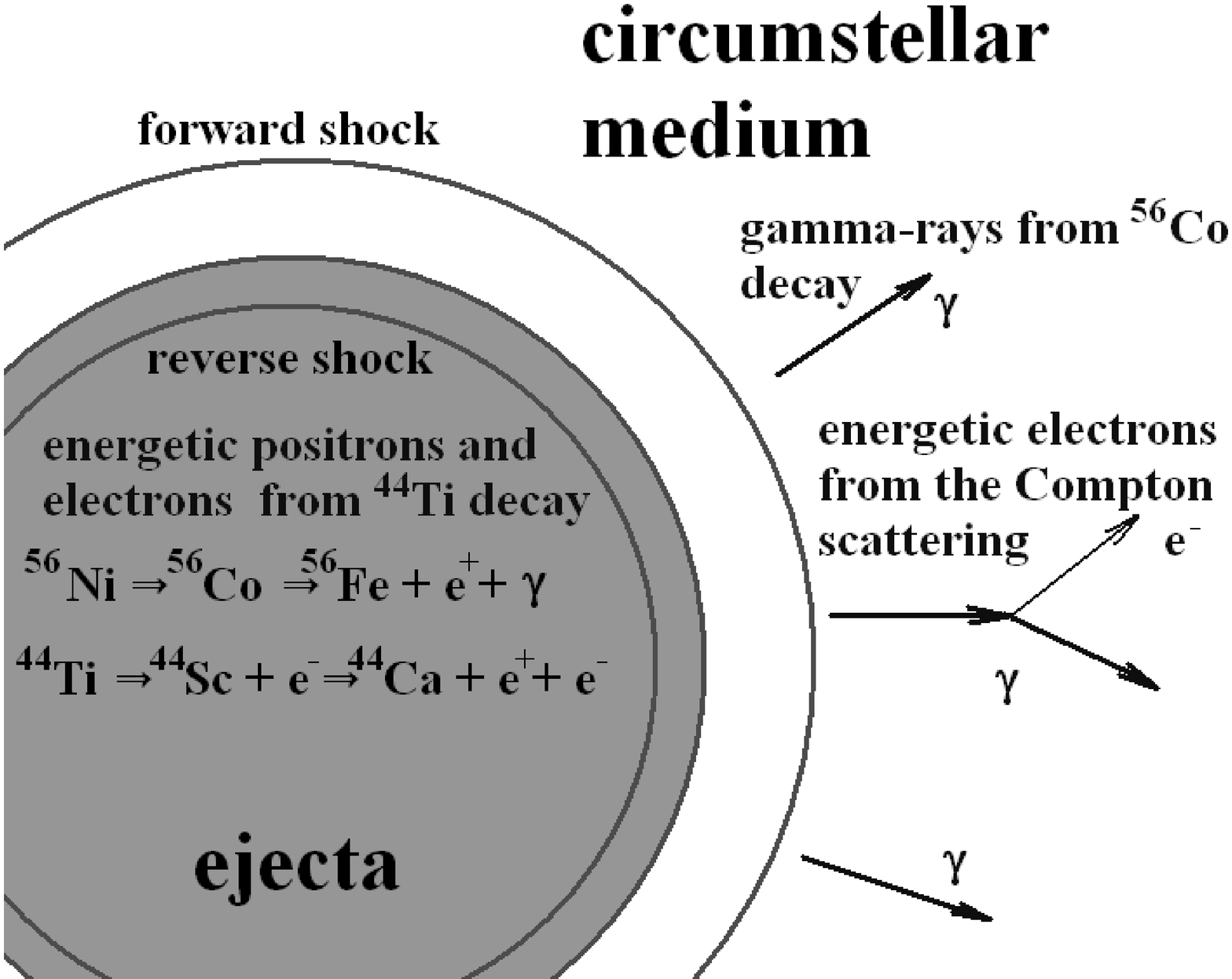}
\caption{Schematic view of a young supernova remnant in the
context of the ``radioactive origin'' of relativistic electrons.
The forward shock propagates in the circumstellar medium outward,
while the  reverse shock  propagates into the ejecta (gray
color) outward in the laboratory frame and inward in the frame of
expanding ejecta. Pre-existing energetic electrons are produced in
the circumstellar medium via the Compton scattering of gamma-rays
from the decay of $^{56}$Co. The  radioactive decays of  $^{44}$Ti
provide energetic electrons and positrons in the ejecta.}
\end{figure}

Cas A, an approximately 300 year old  remnant,
shows bright  broad-band emission extending from radio to  gamma-rays.
It consists of both thermal and nonthermal components, indicating the presence
of hot  thermal plasma, strong magnetic field,  relativistic electrons, and likely also protons,
accelerated up to multi-TeV  energies.
All these components constitute a significant fraction
of the bulk motion kinetic energy of the shell expanding with a
speed of 4000 to 6000 km s$^{-1}$ \cite{patnaude09}. Most likely,
acceleration of electrons takes place  both in forward and reverse shocks.

Thin non-thermal X-ray filaments detected at the periphery of the remnant \cite{gotthelf01}
reveal the presence of a strong $\sim $ 1~mG magnetic field \cite{voelk05} and
multi-TeV electrons accelerated at the forward shock of Cas A.   Synchrotron X-rays are
produced both  in the reverse and  forward shocks \cite{helder08}.  The  time
variations of synchrotron X-radiation  found for a number of
filaments and  knots associated with the reverse shock, indicate that
magnetic field in these compact  structures also is  very large,
close  to 1~mG\cite{Uchi08}.   Because of  large  magnetic fields,
gamma-rays produced via inverse Compton scattering is strongly suppressed,
except  for some regions in the reverse shock with relatively small
magnetic field. Even so, the total energy in protons, assuming that the detected
GeV \cite{abdo10} and  TeV gamma-rays \cite{aharonian01,albert07,acciari10}
are of purely hadronic origin,  does not significantly exceed $10^{49}$erg \cite{abdo10}.
On the other hand,  the  bright  synchrotron
radio emission of Cas~A  indicates to the  existence of  huge amount of relativistic electrons
accelerated by forward and reverse shocks
with total  energy as large as $10^{48} \ \rm erg$ \cite{atoyan00}.
That constitutes approximately  $10^{-3}$ fraction of the  explosion (mechanical) energy.
A significant fraction  of this energy is contained in compact
radio-knots of Cas~A \cite{tuffs86, anderson95}
where  the pressure  of  relativistic  electrons is  comparable to
the thermal pressure  of the shell.

\section{Production of energetic electrons and positrons}

Below  we assume that the radioactive
elements are distributed uniformly throughout the ejecta. Although
these elements are synthesized predominantly
in the core of  the ejecta,  during the  explosion
they can be well  mixed  in  the ejecta.

The ratio of number density of energetic MeV positrons $n_+$ from
$\beta $-decay of $^{44}$Ti to the baryonic density of the ejecta
$n_{ej}$ is given by
\begin{equation}
\frac {n_+}{n_{ej}}=0.94\frac {M_{Ti}}{44M_{ej}}\left[ 1-\exp
\left( -\frac {t\ln 2}{t_{1/2}}\right) \right] .
\end{equation}
Here $t$ is the time since supernova explosion, $M_{ej}$ is
the mass of ejecta, and  $M_{Ti}$  is the  total mass of the ejected nuclei $^{44}$Ti.
Eq. (1)  takes into  account that positrons appear in $94\% $ of the $^{44}$Sc decay.

The rate of Coulomb energy losses of electrons and positrons is described as
\[
\frac{\dot{E}}{E}=\frac {4\pi r_e^2m_ec^4n_{ej}}{vE}\Lambda  \left< \frac ZA\right> =
\frac {r_e^2m_ec^4}{vE}\Lambda \frac {3(k-3)M_{ej}}{2km_pV_{ej}^3t^3}
\sim
\]
\begin{equation}
\frac {0.025}{t}\frac {m_ec^3}{vE}\frac {\Lambda }{40}
\left( \frac {M_{ej}}{M_{\odot }}\right) ^{5/2}
\left( \frac {E_{SN}}{10^{51}\ \mathrm{erg}}\right) ^{-3/2}\left( \frac {t}{63\ \mathrm{yr}}\right) ^{-2}.
\end{equation}
Here $r_e$ is the classical electron radius, $m_p$ and $m_e$ are the proton and electron masses respectively,
$\Lambda \simeq 40$
is the Coulomb logarithm in fully ionized plasma,
$E_{SN}$ is the total energy of explosion, and
$V_{ej} = \left( 10(k-5)E_{SN} /3(k-3)M_{ej}\right) ^{1/2}$ is the characteristic velocity of ejecta
with a power-law  density distribution characterized by the  index $k \sim 10$ \cite{chevalier82b}.
In Eq. (2)  the mean ratio of the atomic number to the mass number $\left< \frac ZA\right> $ is
taken 0.5.
Note  that,  in addition to positrons with energy $E_+\sim $1 MeV,
one electron of  energy $E_-\sim $0.1 MeV is produced  per a $^{44}$Ti decay.
However,  because of difference in energies
the positrons have more chances to be accelerated before they are  thermalized.  Therefore
the  fraction of the  accelerated  positrons  $n_+/(n_++ n_-) \geq 1/2$.

For supernova explosions with small ejecta masses, $M_{ej}<5M_{\odot}$,
the energy losses of positrons from decays of $^{44}$Ti are  not  significant
(see also \cite{martin10}). For  larger ejecta  masses, the energetic positrons
are thermalized  before they
are injected into the reverse shock.  In any case,  these particles  cannot
travel and approach  the forward shock. In this regard, $^{44}$Ti
cannot provide electrons and positrons for acceleration by the
forward shock. Nevertheless,  the forward shock  can be
supplied by  suprathermal  electrons, but  through a different (indirect)  way
related to the Compton scattering of  MeV  gamma-rays - the products of
$^{56}$Co decays.

The number density of energetic  electrons of Compton origin produced by
MeV gamma-rays from $^{56}$Co decays in the
circumstellar medium with number density $n$ is estimated as
\begin{equation}
\frac {n_-}{n}=\xi _{\gamma }\frac {M_{Ni}}{56m_p}\frac {\sigma _{\rm T}}{4\pi r^2}
\sim 1.2\cdot 10^{-7}\xi _{\gamma }\frac {M_{Ni}}{M_{\odot }}r^{-2}_{\rm pc}.
\end{equation}
Here $\sigma _{\rm T}$ is the Thompson cross-section, $r$ is the
distance from the center of the supernova explosion and
$\xi _{\gamma }$ is the fraction of gamma-rays which escape
the expanding ejecta. For photons of  energy of $E \sim 0.5$ MeV
the cross-section of the Compton scattering is
$\sigma _{\rm C}=0.4\sigma _{\rm T}$. It is  taken into account in Eq.(3) that
in a single act of decay of  $^{56}$Co
on average  2.5  gamma-ray photons are produced.
We should note that a similar idea  for the production  of
energetic electrons in SNRs via the Compton scattering of
gamma-rays from the annihilation of $^{56}$Co decay positrons has been
earlier suggested by Bychkov \cite{bychkov77}. This gives
additional  0.5 gamma-photons per a decay of $^{56}$Co.

In the interstellar medium, the timescale of the Coulomb and ionization losses of
energetic electrons  is  of the order of
$10^5$ years. During $300$ years they cannot  diffuse away beyond
3 pc, given that  the diffusion coefficient that characterizes
their propagation does not exceed the standard value of the diffusion coefficient in
the interstellar medium,  $D \sim 10^{28}$ cm s$^{-1}$. Therefore
they will be picked  up by the  arriving  SNR shock.

The fraction of gamma-rays  that escape  the supernova ejecta
is determined by the optical depth $\tau $:
\[
\tau =\left< \frac ZA\right> \sigma _{\rm C}\int n_{ej}dr
=\frac {3(k-3)M_{ej}\sigma _{\rm C}}{4\pi (k-1)m_pV^2_{ej}t^2}\left< \frac ZA\right> \sim
\]
\begin{equation}
0.6\left( \frac {M_{ej}}{M_{\odot }}\right) ^2
\left( \frac {E_{SN}}{10^{51}\ \mathrm{erg}}\right) ^{-1}
\left( \frac {t}{77\mathrm{days}}\right) ^{-2}.
\end{equation}

In order to escape the ejecta without significant lose of energy,
the Compton optical depth for gamma-rays  $\tau$  should not significantly exceed 1.
This determines the time $t$ and the corresponding amount of
non-decaying $^{56}$Co. As it follows from Eq.(4) gamma-rays
from decays of $^{56}$Co can escape the ejecta only if the mass of latter
does not exceed several solar masses. For larger  ejecta masses, the contribution of
gamma-rays from longer-lived
isotopes, e.g.  $^{57}$Co ($t_{1/2}=272$ days, mass $\sim 0.003\ M_{\odot }$ \cite{meyer95}),
becomes more important.

Note that for any reasonable parameters, the Compton optical depth
in the interstellar medium is much smaller than one (even in the
galactic scales), therefore only a small fraction of energy
released at $^{56}$Co decays is transferred to energetic electrons
in the circumstellar medium. The main fraction of energy goes to
the heating of the ejecta.

\section*{Acceleration of electrons}

At the plane non-modified shock with compression ratio $\sigma$,
the far-upstream and downstream momentum distributions of particles,
$F_0(p)$ and $F(p)$, respectively,  are related as

\begin{equation}
F(p)=\gamma \int ^p_0\frac {dp'}{p'}\left( \frac {p'}{p}\right) ^{\gamma }F_0(p').
\end{equation}
Here $\gamma =3\sigma/(\sigma -1)$ is the  Krymsky's index.

Let us assume now that the suprathermal  electrons with a mean energy
$E_{\mathrm{inj}}$ are injected into the plane shock.
For a non-modified strong shock with
compression ratio $\sigma =4$  we have the following expression for
the pressure of accelerated electrons:
\begin{equation}
P_-=\frac 43n_-E_{\mathrm{inj}}\ln \frac {E_{\max }}{E_{\mathrm{inj}}}.
\end{equation}
Here $E_{\max }$ is the maximum energy of electrons accelerated at
the shock. In young SNRs  $E_{\max }$ is of the order of $10-100$ TeV.
Using the number density given by Eq. (1), we can estimate the ratio of the pressure
of positrons $P_+$ to the ram pressure of the reverse shock, $\rho u_r^2$,
propagating  at $t>>t_{1/2}$  into the
ejecta  with a  speed $u_r$:

\[
\frac {P_+}{\rho u_r^2}=\frac {4}{3}\frac {0.94M_{Ti}}{44M_{ej}}
\frac{E_{\mathrm{inj}}}{m_pu_r^2}\ln \frac {E_{\max }}{E_{\mathrm{inj}}}
\sim
\]
\begin{equation}
2.7\frac {M_{Ti}}{M_{ej}}E_{\mathrm{inj}}^{\mathrm{MeV}}
\left( \frac {u_r}{10^3\mathrm{km}\ \mathrm{s}^{-1}}\right) ^{-2}
\ln \frac {E_{\max }}{E_{\mathrm{inj}}}.
\end{equation}
A similar estimate  for the ratio of the electron pressure to the ram pressure $\rho u_f^2$ of the
forward shock propagating in the circumstellar medium with a  speed $u_f$,  gives
\[
\frac {P_-}{\rho u_f^2}=\frac {4}{3}\xi _{\gamma }\frac {M_{Ni}}{56m_{p}}\frac {\sigma _{\rm T}}{4\pi r^2}
\frac{E_{\mathrm{inj}}}{m_pu_f^2}\ln \frac {E_{\max }}{E_{\mathrm{inj}}}\sim
\]
\begin{equation}
1.5\cdot 10^{-5}\xi _{\gamma }\frac {M_{Ni}}{M_{\odot }}E_{\mathrm{inj}}^{\mathrm{MeV}} r^{-2}_{\mathrm{pc}}
\left( \frac {u_f}{10^3\mathrm{km}\ \mathrm{s}^{-1}}\right) ^{-2}
\ln \frac {E_{\max }}{E_{\mathrm{inj}}}.
\end{equation}

From these equations  follows that  the ratio of the electron pressure
to the ram pressure can vary, depending on the several principal
model parameters, within a broad range,  from   $10^{-7}$ to
$10^{-3}$. We  assume   that electrons  are
injected with their original energy $\sim$1 MeV. However their
energy can  be significantly larger  if  particles are
pre-accelerated in the upstream regions of the shocks.

\section{Pre-acceleration of electrons}

High-energy particles
accelerated at strong  shocks excite plasma waves and produce small-scale shocks  and
turbulence in the upstream region. The  turbulence may
amplify magnetic fields at the shocks of  young SNRs \cite{bell04}.
Also, the dissipation of  the turbulence results
in substantial gas heating upstream of the shock. The latter
 limits the total compression ratio of the shock  modified by cosmic rays.  This
is an important feature of  modern nonlinear shock acceleration models (see for a review
ref.\cite{malkov01}).
At  these conditions,  some
pre-acceleration of energetic electrons via the stochastic (second
order Fermi) mechanism  which also energizes thermal electrons and ions
in this region seems  rather  plausible. Note that in principle the stochastic
acceleration can be realized also via  ensemble of random shocks.
Also we should emphasize that there is an essential  difference between
the pre-existing energetic (supra-thermal) electrons and those, which in principle could be
injected at the shock front from the thermal pool. While the pre-existing
energetic electrons pass through the whole extended turbulent
region upstream of the shock,  the particles injected in the
shock front occupy a narrow region  at the shock. That is why
pre-acceleration of these electrons  is  not significant.
The reacceleration of sub-keV electrons from the thermal pool
of upstream plasma is problematic  also
because of strong Coulomb losses (see Eq. (2)).

The energy $E_{\mathrm{inj}}$ is determined by the efficiency of
stochastic acceleration upstream of the shock. The rate of
stochastic (second order) acceleration is $\tau _{st}^{-1}\sim
u_t^2/D$ while the rate of DSA is $\tau _{D}^{-1}\sim u^2/D$,
where $u_t$ is the  the velocity of turbulence (plasma waves) and
$D$ is the diffusion coefficient. The maximum energy of protons is
of the order of 100 TeV in young SNRs. Then for $u_t/u\sim 0.1$,
the maximum energy accelerated through the stochastic mechanism is
expected $E_{\mathrm{inj}}\sim 1$ TeV.  However, this  should be
considered as an optimistic upper limit, given that the diffusion
coefficient  for the low-energy particles in the turbulent region
upstream of the shock can be significantly larger than the Bohm
diffusion.

A more realistic estimate is given below. We shall consider the reacceleration of particles by multiple
 small-scale shocks in the upstream region of the SNR shock. A particle  is picked-up by the
 small-scale shocks, accelerated and advected downstream where it loses energy adiabatically.
 Then  the particle is picked-up by the next small-scale shock,  {\it etc}.

The energy density
 of relativistic electrons just
 downstream of the small-scale shock can be found after integration of Eq. (5).
 Because of the  adiabatic expansion in the
 downstream region,  this value drops by a factor of $\sigma _s^{4/3}$,
 where $\sigma _s$ is the compression
 ratio of the small-scale shock. So  the energy density  $\epsilon _-$ after
 one acceleration cycle is

\begin{equation}
\frac {\epsilon _-}{\epsilon _0}=\frac {\gamma _s}{\gamma _s-4}\sigma _s^{-4/3}=
\frac {3\sigma _s}{4-\sigma _s}\sigma _s^{-4/3}.
\end{equation}
Here $\epsilon _0$ is the electron energy density at the beginning of the cycle.

It is interesting to compare the relative change of the electron energy density to  the relative
change of the gas pressure $P$. Using
the Rankine-Hugoniot conditions we find

\begin{equation}
\frac P{P_0}=\frac {4\sigma _s-1}{4-\sigma _s}\sigma _s^{-5/3}.
\end{equation}
Here $P _0$ is the gas pressure in the beginning of the cycle.

One can see that the relative changes of the electron energy density and of the gas
pressure are similar.  For example, for $\sigma _s=3$ we have
${\epsilon _- }/{\epsilon _0}=2.08$ and ${P}/{P _0}=1.76$. For weaker
shocks, the change of the electron energy density is higher than the change
of the gas pressure. This means that after many cycles,  the relative change of the gas
pressure is comparable or smaller than the change of the electron
energy density. In other words,  the gas heating in the upstream
region of the SNR shock is accompanied by a similar or stronger
electron reacceleration.

Although the gas heating can not directly estimated from observations of SNRs,
one can constrain it (a lower bound)
assuming non-negligible amplification of the magnetic field.
Numerical studies  of the Bell's instability show that the energy
 density of the heated gas is comparable or
 higher than the energy of the amplified magnetic field\cite{bell04,zirakashvili08,riquelme09}.
 Namely,  within the synchrotron-loss interpretation of thin
X-ray filaments in young SNRs (see e.g. ref.\cite{voelk05}), the field  in the upstream region
can  be amplified by a factor of 5 to 10.  Therefore the
gas pressure should be increases by a factor as large as   100.
The similar level of the gas heating is needed
 to limit the strong shock modification
 and to avoid the appearance of the  concave CR spectra (see e.g. ref.\cite{malkov01}).
 It is sufficient  to have 8 cycles to provide
 a 100-fold increase of the gas pressure at the shocks with $\sigma _s=3$. The corresponding increase of
 $E_{inj}=\epsilon _-/n_-$ equals several hundreds.

The modeling of the Bell's instability
 with DSA \cite{zirakashvili08} shows  that the upstream region of a young SNR of width $L\sim 10^{18}$ cm
 is filled with a supersonic MHD turbulence with Mach number 3-4, while  the distance between small-scale
 shocks is $l\sim 10^{16}$ cm. For these parameters and for turbulent motions
$u_t/u\sim 0.1$ the expected number of cycles is $Lu_t/lu\sim 10$.

One should note that the
 pre-accelerated electrons  may have an impact  on the upstream
turbulence and thus regulate their own acceleration efficiency. In
particular,  the higher number density of pre-existing electrons would make
lower the energy $E_{\mathrm{inj}}$. Under these conditions,  the
energy density of pre-accelerated electrons may be of the order of
the energy density of the upstream turbulence. The latter is believed to be several
percent of the ram pressure $\rho u^2$ at cosmic ray modified  shocks. So the upper limit for the
number density of pre-accelerated  electrons is $n_-E_{\mathrm{inj}}\sim 10^{-2}\rho u^2$
in Eq. (6).

Even for a modest
energy  $E_{\mathrm{inj}}=100$ MeV,  one can obtain, according to Eq.(7),
quite high ratio $P_+/\rho u_r^2\sim 0.1$.
The shock may be slightly  modified by the pressure of accelerated
electrons and positrons!

\section{Applications to SNRs}

The above discussed picture of pre-acceleration of electrons and
positrons from  the products of decays of radioactive short-lived
elements can be relevant to the  reverse shock of Cas A. This can
explain why the pressure of energetic positrons (electrons) in the
shocked ejecta is  comparable to the gas pressure in the supernova
shell. The same could be true also for the radio-knots if they are
fast moving clumps of the shocked ejecta. At the present epoch,
the  pressure of energetic electrons at the forward shock of Cas~A
is not very high,   as it follows from Eq.(8). However, most
likely  it was much higher  in the past when the radius of the
remnant was smaller than 0.1 pc. Since the forward shock of Cas~A
propagates in a dense stellar wind of the supernova progenitor
with a density profile $\sim r^{-2}$, the accelerated electrons
have been produced mainly in the past when the synchrotron cooling
in the amplified field was significant. Now these electrons are
located inside the forward shock. This  can explain the  rather
steep radio-spectrum  of Cas A.

In Cas~A,   the energy of pre-accelerated electrons $E_{\rm inj}$
can not exceed  100-200 MeV, otherwise this would  be in conflict
with the observed  synchrotron radio spectrum.
The  spectral flatening seen at 20 MHz \cite{baars77}
can  be attributed,  for the magnetic field at the reverse
shock of the order of 100-200 $\mu $G, to the lower energy cut-off in the
electron spectrum at 100 MeV.  The magnetic
field at the forward shock of  Cas A is  larger.
However since radio-emitting electrons have been  accelerated
in this region in the past,  because of adiabatic losses
their low-energy cut-off is now located below 100 MeV

We conclude that the high radio brightness of Cas~A
is caused  by  the dense stellar wind where the
forward shock propagates, and by  a relatively high amount of radioactive
$^{44}$Ti  the decay of which
provides supra-thermal electrons and positrons for the  further
acceleration  by  the reverse shock. This is in contrast to  other historical
young SNRs like Tycho, Kepler and SN1006. They are results of Ia supernova explosions
in  uniform medium.  Therefore, in these objects the electrons
accelerated  by forward shocks, are  produced predominantly at later epochs.
In addition, the ejecta of Cas~A, because of the dense stellar wind  has been shocked
very early,  likely just after the explosion. The radiative instabilities
operated in the shocked ejecta, could  result in the  formation of ejecta
clumps \cite{hwang03,hwang09}, which presently are observed as radio-knots.

Since the reverse shock of Cas~A contains about 1$\%$ of  the
explosion energy, the energy fraction of electrons and
positrons is close to $10^{-3}$. The electrons accelerated at the forward shock have a similar energetics. So
 we expect that in Cas~A  approximately
 $10^{-3}$ fraction of supernova energy is transferred to the accelerated
electrons and positrons.  This conclusion is in agreement with estimates
based on radio observations \cite{atoyan00}.

The fraction of energy $10^{-3}$ found for positrons in the reverse shock of
Cas~A is expected to be
the same for all young core-collapse supernova. However GeV
positrons  leave the remnant only at late stages when its
radius becomes  a factor of 10 larger than the radius at the transition
to the Sedov phase when the positrons have been accelerated. Since the energy of
particles adiabatically drops (inverse proportional to the
remnant's radius), the energy fraction of positrons will  be reduced down
to $10^{-4}$. The
luminosity in galactic CR positrons at multi-GeV energies
based on the recent measurements of the Pamela collaboration \cite{adriani09} is close to
$10^{38}$ erg s$^{-1}$. Given the overall  mechanical power of the galactic core
collapse supernova  $10^{42}$ erg s$^{-1}$, our model can explain the flux
of the primary CR positrons by reverse shocks of young SNRs
without invoking other source populations (for
a review on different potential sources of galactic CR positrons see \cite{Positrons}).

It is important to note that our model applied to Cas~A  predicts
the positron-to-electron ratio close to 1. The reason is that (i)
the estimated energetics of leptons in forward and reverse shocks
in  Cas A based on radio observations are comparable, and (ii) our
model implies that while electrons are accelerated in the forward
shock, in the reverse shock the content of positrons is equal or
larger than the content of electrons.  If so, Cas A, as well as other young SNRs
alone cannot provide the total flux of galactic CR
electrons. In fact this is a model-independent statement based on
the estimates of numbers of electrons in young SNRs.  For old SNRs
the situation is  different. While the reverse shocks disappear in
these objects, the forward shock continue to accelerate electrons
(although to modest energies, $E \leq 1$~TeV). In our model, the
electrons produced via the Compton scattering of gamma-rays from
$^{56}$Co are accelerated by  forward shocks in  old Ia SNRs
expanding in the uniform medium.  According to the observed light
curves, the ejecta of Ia supernova contains $\sim $ 0.6$M_{\odot }$
of $^{56}$Ni just after the supernova explosion.  The energetics
of galactic Ia supernova is of the order of $3\cdot 10^{41}$ erg
s$^{-1}$, implying approximately  one supernova per century. On
the other hand, the  production rate of galactic CR
electrons is close to $10^{39}$ erg s$^{-1}$\cite{berezinsky90}.
So a  fraction of $0.3\% $  of energy of Ia supernova must  be
transferred to CR electrons. The similar ratio of cosmic
ray electron pressure to the ram pressure  is estimated for an old
remnant with the radius $30$ pc and the shock speed  $300$ km
s$^{-1}$ if $E_{\mathrm{inj}}=3$GeV (see Eq. (9)). The required  higher
value of $E_{\mathrm{inj}}$ can be explained by  a lower number density of
the circumstellar medium where  the Ia supernova explosions occur.

We should note that another source of the suprathermal electrons
at supernova shocks has been recently  suggested  by Morlino
\cite{morlino11}. Partially ionized
 multi-GeV ions accelerated at the shock can produce multi-MeV electrons via
 photo-ionization by optical Galactic emission. The fraction $\eta =n_-/n$
 of the corresponding electrons is
 estimated  as $\eta \sim 0.1x_{He}\gamma ^{-1}u^2/c^2$ at cosmic-ray modified shocks.
 Here $x_{He}\sim 0.1$ is the fraction of Helium in the interstellar medium,
 $\gamma \sim I_{He}/\epsilon _{ph}$ is the gamma-factor of He$^{+}$ ion ionized by Galactic
 optical photons with energy $\epsilon _{ph}$, and $I_{He}=54$ eV is the ionization
potential of Helium. This results in $\eta \sim 10^{-4}u^2/c^2$ in young SNRs where $\gamma \sim 100$
 and ions are photo-ionized by eV optical photons, and $\eta \sim 10^{-3}u^2/c^2$ in the
 old remnants where $\gamma \sim 10$ and ions are photo-ionized by ultraviolet photons. These numbers
 are comparable or higher than numbers given by Eq. (3).
 Even without any preacceleration by MHD turbulence this mechanism
 results in the electron to proton ratio $K_{ep}\sim x_{He}m/m_p\sim 10^{-4}$. Although the preacceleration of
 these electrons is more problematic because they are produced closer to the shock by 10-100 GeV ions, it is not
 excluded. Then the corresponding injection energy necessary for explanation of galactic cosmic ray
 electrons can  be below 1 GeV closer to $E_{inj}=100$ MeV  as argued above
 for reverse shock of Cas A.

Finally, in the context of the proposed model,
one can expect harder CR positron spectrum. The positrons
of higher energies leave the remnant earlier and are subject to
lower adiabatic losses in comparison with the positrons of lower energies.
This effect does not have an impact  on the spectra of
electrons  accelerated  predominantly
by forward shocks in  old SNRs.
The harder source spectrum
of positrons is in agreement with the recent Pamela measurements \cite{adriani09}.

According to the scenario proposed in this paper, only forward
shocks of young SNRs produced by supernova explosions with a small
ejecta masses $M_{ej}<2M_{\odot }$, can contain large amount of
accelerated electrons. The relevant  SNRs belong to  the
Ia/b/c and, probably,  IIb (like Cas~A)  type supernovae. Note
that the  brightest in TeV gamma-rays young SNR RX J1713.7-3946
most likely belongs to Ib/c type SNR with a  small ejecta mass
\cite{zirakashvili10a}. In the case of {\bf IIP}
supernova with large ejecta masses   gamma-rays from $^{56}$Co
decay  cannot  effectively escape the ejecta and  "feed" the
forward shock by suprathermal electrons for further acceleration.
If so, we should expect forward shocks of IIP SNRs to be  dim in radio and
non-thermal X-rays. On the other hand,  large amount  of electrons
and positrons from decays of $^{44}$Ti  can be accelerated at reverse
shocks  of young SNRs of all types including the most frequent IIP
supernovae. In this regard, the youngest
galactic SNR  G1.9+0.3 is  of a special interest. It shows both large content of
$^{44}$Ti and  ongoing acceleration of electrons by reverse shock
\cite{borkowski10} - two key components  required in  our model.

\section{Summary}

 The "radioactive"  origin  of electron injection, related to both the  forward and
reverse shocks,  seems to be  a natural scenario in SNRs with the following key  components:

1) the energetic positrons (and possibly also electrons) from $^{44}$Ti decay are
accelerated at  reverse shocks of young SNRs;

2) the energetic electrons from the Compton scattering of
$^{56}$Co-decay gamma-rays are accelerated at forward shocks of both old and young SNRs
of type Ia/b/c and IIb;

3) a modest  pre-acceleration (presumably of stochastic origin) to
energies $E_{\mathrm{inj}}\sim 0.1$ GeV in the upstream regions of  the
forward and reverse shocks is a necessary condition in Cas A  for explanation
of  the energetics in  relativistic electrons;

4) the  proposed scenario can explain not only the overall flux of galactic CR electrons
by SNRs, but also the  recently reported tendency of gradual increase of the
positron-to-electron ratio
with energy.







\end{document}